%% file: main.tex
\documentclass[11pt]{article}

\usepackage[a4paper,margin=1in]{geometry}
\usepackage{amsmath,amssymb,amsfonts}
\usepackage{graphicx}
\usepackage{booktabs}
\usepackage{hyperref}
\usepackage{natbib}

\title{%
A Formal Descriptive Language for Learning Dynamics\\[0.5ex]
\large A Five-Layer Structural Coordinate System
}

\author{%
Miyuki T. Nakata\thanks{mtnakata@kumamoto-u.ac.jp} \\
\small Faculty of Advanced Science and Technology, Kumamoto University
}

\date{} 

\begin{document}

\maketitle

\begin{abstract}

Understanding learning as a dynamic process is challenging due to the interaction of multiple factors, including cognitive load, internal state change, and subjective evaluation. Existing approaches often address these elements in isolation, limiting the ability to describe learning phenomena within a unified and structurally explicit framework.

This paper proposes a multi-layer formal descriptive framework for learning dynamics. Rather than offering a predictive or prescriptive model, the framework introduces a symbolic language composed of state variables, mappings, and layer-specific responsibilities, enabling consistent description of learning processes without commitment to specific functional forms or optimization objectives.

This descriptive framework is intended to serve as a structural substrate for analyzing learning processes in human learners, and by extension, in adaptive and Al-assisted learning systems.

A central design principle is the explicit separation of descriptive responsibilities across layers, distinguishing load generation, internal understanding transformation, observation, and evaluation. Within this structure, cognitive load is treated as a relational quantity arising from interactions between external input and internal organization, while subjective evaluation is modeled as a minimal regulatory interface responding to learning dynamics and environmental conditions.

By emphasizing descriptive clarity and extensibility, the framework provides a common language for organizing existing theories and supporting future empirical and theoretical work.

\end{abstract}

\input{sections/s01_introduction}
\input{sections/s02_related_work}
\input{sections/s03_framework_definition}
\input{sections/s04_layer_responsibilities}
\input{sections/s05_positioning}
\input{sections/s06_conclusions}
\input{sections/s07_future_work}

\appendix
\input{appendices/appA_structural_stability}

\section*{Acknowledgments}
\input{sections/s08_acknowledgements}

\bibliographystyle{plainnat}
\bibliography{refs}

\end{document}

%% file: sections/s01_introduction.tex
\section{Introduction}

\subsection{Motivation}

Learning is commonly described using fragmented perspectives: instructional design focuses on task structure, cognitive theories emphasize internal capacity limits, and motivational theories address persistence and disengagement. While each perspective captures important aspects of learning, they often operate with incompatible assumptions, vocabularies, and levels of abstraction.

As a result, it remains difficult to describe learning phenomena such as acceleration, stagnation, or withdrawal in a unified and structurally explicit manner, particularly without attributing outcomes directly to fixed personal traits such as ability or motivation.

This work is motivated by a simple but persistent question:

\begin{quote}
Can learning dynamics be described using a formal language that makes structural dependencies explicit, without committing to a specific predictive model?
\end{quote}

Rather than proposing a new learning theory or algorithm, we aim to construct a descriptive symbolic framework that clarifies where and how different processes intervene in learning.

\subsection{Our Approach: A Formal Descriptive Language}

We propose a multi-layer formal framework designed explicitly as a language for describing learning dynamics.

The framework introduces:
\begin{itemize}
  \item State variables representing internal understanding and generalizable capability
  \item Mappings that describe transformations between descriptively distinct layers
  \item A layered causal structure that enforces explicit separation of responsibilities
\end{itemize}

Crucially, the framework is not intended to predict numerical outcomes. No specific functional forms are imposed, and no optimization criteria are assumed.

Instead, the goal is to provide a structural vocabulary that allows researchers and practitioners to:
\begin{itemize}
  \item Distinguish between different sources of learning difficulty
  \item Identify where interventions act within the learning process
  \item Describe learning trajectories without reducing them to fixed traits or global explanations
\end{itemize}

In this sense, the framework functions analogously to a programming language: it defines permissible structures and relations, as well as constraints on their use, rather than specific programs or solutions.

\subsection{Separation of Concerns as a Design Principle}

A central design principle of the proposed framework is the explicit separation of descriptive responsibilities across the entire learning process.

Rather than treating learning as a single explanatory mechanism, the framework decomposes it into distinct layers, each responsible for a specific descriptive role: the existence of external input, its decomposition relative to representational bases, the internal transformation of understanding over time, the externalization of internal state, and the modulation of the learning loop through subjective evaluation.

Within this structure, processes such as load generation and understanding dynamics are assigned to separate mappings with non-overlapping responsibilities. More generally, no layer is permitted to explain phenomena outside its designated scope. In particular, the framework prohibits interpreting load generation as learning, internal state change as evaluation, or persistence and withdrawal as properties of fixed learner traits.

By enforcing this separation, the framework avoids circular definitions and makes theoretical commitments explicit. It also enables existing theories, such as Cognitive Load Theory, to be situated as partial specifications operating within particular layers or mappings, rather than as competing end-to-end explanations of learning.

This principle of separation is not merely methodological, but structural: it defines what kinds of statements are admissible at each level of description, thereby providing a common language in which heterogeneous theories and observations can be related without conceptual collapse.

\subsection{Limitations of Existing Approaches}

Several influential frameworks address parts of the learning process, yet each leaves structural gaps when considered in isolation.

Cognitive Load Theory (CLT) provides a powerful vocabulary for discussing limitations of working memory and instructional design. However, its focus lies primarily on task-level load categorization and optimization, offering limited means to describe how internal understanding states evolve over time.

Motivation theories, such as Self-Determination Theory, richly characterize psychological needs and motivational orientations, but often rely on latent constructs that are difficult to connect to observable learning dynamics or concrete intervention points.

Reinforcement learning models formalize reward-driven behavior with mathematical rigor, yet typically abstract away the internal representational changes that constitute understanding.

These approaches are not incorrect; rather, they operate at different descriptive layers of the learning process. What is missing is a shared descriptive language that can express how these layers relate and interact, without collapsing them into a single explanatory dimension. The issue is not the inadequacy of these approaches, but the absence of a common descriptive layer that makes their relationships explicit.

\subsection{Scope and Contribution}

The contributions of this work are:
\begin{enumerate}
  \item A formally defined, multi-layer descriptive framework for learning dynamics
  \item An explicit separation of descriptive responsibilities across external input, load generation, internal state transformation, observation, and subjective evaluation
  \item A principled approach to leaving key components deliberately unspecified, preserving extensibility and cross-theoretical compatibility
  \item A positioning of existing learning theories as expressible instances within a shared structural coordinate system, rather than as competing end-to-end explanations
\end{enumerate}

Rather than proposing a new learning theory, the present work introduces a formal descriptive language that constrains how learning processes can be articulated, compared, and related, without committing to predictive models, optimization objectives, or normative assumptions.

%% file: sections/s02_related_work.tex
\section{Related Work}

This chapter positions existing learning theories within the proposed framework by clarifying the descriptive roles they assign to different components of the learning process.

\subsection{Cognitive Load Theory and Instructional Design}

Cognitive Load Theory (CLT) has provided a highly influential set of principles for instructional design by foregrounding the limitations of working memory and by distinguishing among intrinsic, extraneous, and germane cognitive load \citep{Sweller1988CognitiveLoad,Sweller2019CognitiveArchitecture}. Within CLT, these distinctions function primarily as design-relevant categories, offering guidance on how instructional materials and task representations should be structured in order to reduce unnecessary processing demands and to support learning-relevant activity. Importantly, later formulations of the theory emphasize that intrinsic load is not a fixed property of instructional materials themselves, but depends on the interaction between task structure and the learner's prior knowledge, while extraneous and germane load are defined in terms of how information is presented and processed rather than as inherent attributes of stimuli.

The present framework adopts CLT's core constraint-based intuition, namely that learning is bounded by limited cognitive resources, while shifting the analytical focus from load categories as classificatory labels to the mechanisms by which load is generated. In this framework, load is treated as the output of explicit load-generation mappings that decompose a given external input relative to available representational bases. These mappings are not modified by the learner's internal state. The state does not select or control load-generation mappings; it only defines the conditions under which a given decomposition is descriptively meaningful at a given time. As a result, intrinsic, extraneous, and germane load are not assumed to reside in instructional materials themselves, but emerge relationally through the interaction between input representations and the learner's current conceptual and structural organization. This separation allows the framework to distinguish analytically between (i) how different kinds of load are induced at the level of load generation ($\Phi$), and (ii) how that induced load is transformed over time into changes in understanding via internal dynamics ($G$). From this perspective, CLT can be situated within the framework as a collection of empirically grounded constraints and design heuristics on particular load-generation mappings, rather than as a complete dynamical account of learning processes.

\subsection{Guidance, Structure, and Failure as a Learning Condition}

Debates on instructional guidance and the literature on productive failure jointly illustrate that learning cannot be characterized as a monotone function of ease, guidance, or observable success. The critique of minimal guidance emphasizes that unguided or weakly guided environments can impose excessive cognitive load on novice learners, often consuming limited working-memory resources without yielding durable learning \citep{Kirschner2006MinimalGuidance}. In contrast, the productive failure line of work has demonstrated that initial struggle or failure can, under certain conditions, contribute positively to deeper learning outcomes \citep{Kapur2016ProductiveFailure}. Subsequent studies have further clarified that such benefits are highly conditional, depending on factors such as task structure, opportunities for consolidation, and the learner's interaction with the environment \citep{SinhaKapur2019PF}.

From the viewpoint of the present framework, these seemingly divergent findings point to the same underlying distinction: learning outcomes are determined not by whether learners struggle or succeed at a given moment, but by how the load induced by an activity is generated and subsequently transformed over time. Unguided or exploratory environments may generate substantial load at the level of load generation ($\Phi$), yet whether this load contributes to learning depends on its alignment with available conceptual and structural bases and on its propagation through the internal dynamics ($G$). In some cases, initial failure gives rise to load that is later reorganized and integrated, supporting subsequent changes in understanding; in other cases, the same apparent failure primarily generates residual load that dissipates without productive transformation, leading to stagnation or disengagement. By explicitly separating load generation from understanding transformation, the framework provides a descriptive vocabulary for articulating these distinctions without assigning normative value to success or failure themselves, locating their effects instead within different structural roles in the learning process.

\subsection{Motivation and Learner Experience}

Motivational theories such as Self-Determination Theory (SDT) provide a rich account of autonomy, competence, and relatedness as determinants of sustained engagement \citep{DeciRyan2000SDT}. These constructs capture essential aspects of learner experience, but they are often introduced as explanatory variables at a different descriptive level than the time evolution of understanding or capability. As a result, motivational states can become ``global labels'' that obscure where, structurally, a learning process becomes self-reinforcing or collapses.

The present framework takes a deliberately different stance in the present formulation: it introduces a minimal subjective reward variable as an interface that responds to internal dynamics, while postponing any decomposition into motivational sub-constructs. This design choice is not a denial of motivational phenomena, but an attempt to preserve a non-attributional structural description in which persistence, stagnation, and withdrawal can be discussed in terms of interactions among load extraction, processing dynamics, observation, and capability growth. SDT and related accounts can then be positioned as higher-level interpretations or partial specifications of how subjective evaluation is shaped by the learning process and its environment.

\subsection{Dynamical Systems Approaches to Cognition and Development}

Dynamical systems perspectives have long argued that cognition and development should be understood as time-evolving processes characterized by non-linearity, multi-timescale interactions, and emergent stability \citep{ThelenSmith1994DynamicSystems,vanGelder1998DynamicalHypothesis}. These approaches are conceptually aligned with the present work in treating learning as dynamics rather than as a static state.

A key difference is methodological orientation. Many dynamical models aim to capture particular phenomena through specific parameterizations and functional forms, often tailored to a domain and an experimental paradigm. The present framework adopts a pre-modeling stance: it introduces state variables, mappings, and a layered causal structure without committing to functional forms. Its objective is not to predict trajectories but to clarify where dynamics occur and where different explanatory traditions place their mechanisms. In this sense, the framework is compatible with dynamical modeling while remaining agnostic about mathematical instantiation.

\subsection{Learning Sciences as a Reference Frame and Positioning}

Contributions from the learning sciences provide an integrative reference frame that spans cognition, instruction, and context, emphasizing that learning outcomes arise from interactions among prior knowledge, activity, environment, and social organization \citep{Bransford2000HowPeopleLearn}. A central message of this tradition is that learning with understanding cannot be reduced to properties of tasks or instructional materials alone, but must be analyzed in relation to the learner's existing knowledge structures and ongoing activity.

At the same time, this integrative strength comes with a descriptive challenge: explanations often operate across multiple levels, including task design, learner cognition, observation, assessment, and motivation, without an explicit separation of their structural roles. As a result, distinct mechanisms may be discussed within a shared narrative, even when they belong to different parts of the causal process.

The present work positions itself not as an alternative comprehensive theory of learning, but as a formal descriptive layer that clarifies this structure. Specifically, it makes explicit (i) how load is generated from external input relative to available representational bases, (ii) how changes in understanding are represented as internal dynamics over time, (iii) how content-related understanding and structural capability are distinguished within a unified state space, and (iv) how observation and subjective evaluation can be located within the same causal graph without being reduced to fixed learner traits. In this way, established learning-sciences theories can be expressed, compared, and related within a shared structural vocabulary, while preserving openness to diverse empirical and theoretical commitments.

%% file: sections/s03_framework_definition.tex
\section{Framework Definition}

This section introduces the formal definitions of the proposed multi-layer descriptive framework.
Each layer is defined by its responsibility and associated mappings, without imposing specific functional forms or normative assumptions.

\subsection{Layer 0: External Input}

Layer~0 represents the external input $e$ available to the learner, encompassing any form of stimulus, signal, or environmental interaction. This may include instructional materials, tasks, perceptual stimuli, social signals, or interactions with artifacts, but is not restricted to educational settings. These examples are illustrative only and do not imply any semantic or functional interpretation at this layer.

The input $e$ is treated abstractly as an element of a representational space $\mathcal{E}$:
\[
e \in \mathcal{E}
\]

No assumptions are made regarding the dimensionality, numeric field, topology, or internal structure of $\mathcal{E}$; in particular, $e$ is not assumed to be a vector in $\mathbb{R}^n$. It may correspond to symbolic, perceptual, relational, temporal, or composite structures, depending on the context in which it arises.

At this layer, $e$ is not attributed any intrinsic meaning, pedagogical intent, or evaluative status. Layer~0 specifies only the existence of an external input available for subsequent processing. All questions concerning how $e$ is decomposed, represented, interpreted, transformed, or evaluated are explicitly deferred to higher layers of the framework.

\subsection{Layer 1: Load Generation}

Layer~1 specifies how external input is decomposed relative to a selected representational basis. Its role is descriptive: it specifies how an input is partitioned with respect to a chosen context, without introducing evaluative or normative assumptions.

Importantly, load generation does not implement learning, interpretation, or state change. Instead, it provides a time-indexed decomposition of the input that serves as an interface between external input and internal transformation processes.

\subsubsection{Contextual bases and representational selection}

Let $x$ denote the learner's internal state, representing the totality of currently available internal organization. The internal state does not modify the load-generation mappings themselves. Rather, it provides conditions under which certain contextual bases may become relevant or operative at a given time, without specifying the mechanism by which such bases are selected.

We consider a family of load-generation mappings $\{\Phi^{(c)}\}$, where each $\Phi^{(c)}$ corresponds to a particular contextual or conceptual basis $c$. Each mapping specifies a representational decomposition of the same external input $e$ relative to that basis.

Crucially, $\Phi^{(c)}$ is not a learning operator. It does not alter $e$, introduce interpretation, or induce internal change. Its sole function is to specify a coordinate system, or more abstractly, a decomposition rule, relative to which load quantities are defined.

The selection of a particular contextual basis $c$ from the family $\{\Phi^{(c)}\}$ is not specified by the present framework. While the accessibility of contextual bases is constrained by the learner's internal state, the actual selection or switching among available mappings may depend on factors outside the scope of Layer~1. Such selection may be viewed as a time-dependent process, potentially exhibiting stochastic characteristics, without being attributed to an explicit internal control variable or learning operation.

\subsubsection{Definition}

For a given contextual basis $c$, the load-generation mapping is defined as:
\[
\Phi^{(c)} : e \mapsto (\ell_x, \ell_n)
\]

where:
\[
\ell_x = \left\| \mathrm{Proj}^{(c)}(e) \right\|
\]
\[
\ell_n = \left\| e - \mathrm{Proj}^{(c)}(e) \right\|
\]

The notation $|\cdot|$ denotes a generalized magnitude measure and does not imply a norm in a strict mathematical sense unless explicitly specified.

Here:
\begin{itemize}
  \item $\mathrm{Proj}^{(c)}$ denotes an abstract projection or extraction operator associated with the contextual basis $c$. The term ``projection'' is used in a generalized sense and does not imply linearity or orthogonality unless otherwise specified.
  \item $\ell_x$ represents the magnitude of the component of the input that is expressible within the selected representational basis.
  \item $\ell_n$ represents the magnitude of the residual component not captured by that basis.
\end{itemize}

The quantities $\ell_x$ and $\ell_n$ are purely descriptive and non-normative. They do not encode correctness, relevance, difficulty, or educational value.

\subsubsection{Interpretation}

Within a given context $c$, $\ell_x$ denotes the portion of the input that is representable under the selected basis. This component is eligible to contribute positively to subsequent internal transformation, as described in Layer~2.

By contrast, $\ell_n$ denotes the portion of the input that is not representable under the current contextual basis. Importantly, $\ell_n$ does not imply noise, error, or irrelevance in a global sense. It is defined strictly relative to the chosen context: a component that appears as residual under one mapping $\Phi^{(c)}$ may become aligned and contribute to $\ell_x$ under a different mapping $\Phi^{(c')}$.

Layer~1 makes no assumptions about whether $\ell_n$ will inhibit, attenuate, or remain neutral with respect to learning. Such effects are not specified at this layer and are instead determined by the internal dynamics described in Layer~2.

\subsubsection{Scope of Layer 1}

No state update occurs within Layer~1. Load generation is evaluated pointwise in time and specifies only a decomposition of the external input. All changes to the internal state $x$, including learning, consolidation, or reorganization, are exclusively described in Layer~2 via the dynamics $G$.

\subsection{Layer 2: Understanding Transformation}

Layer~2 describes how generated load is transformed into temporal changes in the learner's internal state. This layer constitutes the sole locus of learning in the proposed framework, where all updates to the internal state occur.

In contrast to approaches that treat content understanding and general capability as separate or independently optimized subsystems, the present framework models them as jointly evolving aspects of a single internal state. Learning is thus described as a dynamical process driven by the interaction between the current state and the structure of the generated load.

\subsubsection{State Variables}

The learner's internal state is represented as:
\[
x(t) = (p(t), f(t))
\]

where:
\begin{itemize}
  \item $p(t)$ denotes context- and content-dependent conceptual organization, reflecting which distinctions, representations, or interpretations are currently stabilized.
  \item $f(t)$ denotes more context-invariant structural properties of the internal dynamics, such as abstraction or relational sensitivity, that support sustained state transformation across varying representational conditions.
\end{itemize}

Formally, within Layer~2, $(p,f)$ are represented as components of a single state space rather than as separable variables. While content-specific organization and structural capability may influence one another over time, no assumption is made that they evolve synchronously or under identical conditions.

To provide intuition, consider a learner repeatedly engaging with tasks from a fixed domain. Observable performance (Layer~3) may remain unchanged, while internal organization improves; for example, solutions may become more compact, representations more internally flexible, or alternative interpretations more stably expressible within the learner’s internal organization. Within the present framework, such changes correspond to evolution in the joint state $(p,f)$, even in the absence of immediate performance gains. This motivates treating content-related and structural aspects as coupled components of a unified internal state.

Accordingly, $p$ tends to depend strongly on the representational basis through which load is generated, whereas $f$ captures aspects of internal dynamics that remain invariant across representational conditions.

\subsubsection{Dynamics}

The temporal evolution of the internal state is described by:
\[
\frac{dx}{dt} = G(x, \ell_x, \ell_n)
\]

Here, $G$ denotes the understanding transformation dynamics determined by the interaction between:
\begin{itemize}
  \item the current internal state $x(t)$
  \item the context-aligned load component $\ell_x$, extracted by the selected mapping $\Phi$
  \item the residual load component $\ell_n$, representing input dimensions not aligned with the current contextual basis
\end{itemize}

The component $\ell_x$ represents input structure that is compatible with the currently selected representational basis and may contribute positively to state change. By contrast, $\ell_n$ represents dimensions of the input that are not captured under the current context. Within Layer~2, $\ell_n$ is not treated as noise or error, but as a component that may have no effect on learning dynamics or may attenuate effective transformation under the current mapping.

The function $G$ specifies how these components jointly influence the evolution of $p$ and $f$ over time. No assumptions are made regarding linearity, monotonicity, optimality, or specific functional form. In particular, increases in $p$ may occur without corresponding growth in $f$, and structural capability $f$ may exhibit saturation or threshold-like behavior depending on the learner's history and the diversity of experienced contexts.

Crucially, Layer~2 does not perform evaluation. It does not determine whether a state change is desirable, successful, or correct. It specifies only how the internal state evolves in response to the generated load. All evaluation, regulation, or modulation of learning trajectories is handled separately in higher layers of the framework.

\subsection{Layer 3: Externalization (Observation)}

Layer~3 maps the internal state to observable quantities without modifying the state itself.
\[
y = Q(x)
\]

Here,
\[
x = (p, f)
\]
denotes the internal state vector comprising both content-related and structural components. The mapping $Q$ represents processes of observation or externalization, such as performance outcomes, verbal explanations, written artifacts, or other measurable behavioral traces, without implying evaluative judgment.

The framework does not assume that all components of $x$ are equally observable. In general, $Q$ is a partial projection that may be indirect, noisy, or incomplete.

\subsection{Layer 4: Subjective Evaluation Interface}

Layer~4 generates an internal signal that modulates the strength of the learning feedback loop, based on the temporal dynamics of the internal state and the external environmental context.
\[
r = R\left(\frac{dx}{dt},\, E\right)
\]

Here, $E$ denotes aspects of the external environment that are not encoded in the internal state itself, such as environmental stability, risk, resource availability, or external constraints on exploration, including feedback arising from the learner's own externalized actions or expressions, whether or not such feedback carries explicit evaluative content, once encountered as environmental input.

The output $r$ represents an internal gain signal that regulates how strongly the learning process couples to subsequent exploration, engagement, or attenuation. Within the proposed framework, $R$ is positioned as an abstract evaluation interface rather than as a reward function in the sense of optimization or reinforcement learning. No specific functional form, utility interpretation, or decision-making mechanism is assumed.

Crucially, $r$ does not depend directly on the current internal state $x$. Instead, it responds to how the internal state is changing and to conditions external to the learner. This design prevents evaluation from collapsing into a value judgment over states and preserves a clear separation between state dynamics (Layer~2) and loop modulation (Layer~4).

A central design principle of this layer is the absence of normative assumptions. The framework does not presuppose that learning persistence is always desirable, nor that disengagement or attenuation constitutes failure. Rather, $r$ is treated as a regulatory signal that modulates the learning loop in response to the interaction between internal change and environmental conditions.

Importantly, the framework distinguishes \emph{evaluation} from value judgment. The signal $r$ does not encode correctness, desirability, or educational success. It specifies only how strongly ongoing internal dynamics are coupled to continued interaction with the environment at a given time.

This formulation allows learning persistence, stagnation, and attenuation to be described as structurally coherent control responses, without attributing them to fixed learner traits or global motivational states. Which regime becomes dominant is left deliberately unspecified at the level of the framework.

By treating subjective evaluation as an interface rather than as a model of motivation, Layer~4 remains compatible with diverse theoretical interpretations, including psychological, biological, and dynamical perspectives, while preserving the framework's role as a non-prescriptive descriptive language.

\begin{figure}[t]
  \centering
  \includegraphics[width=\linewidth]{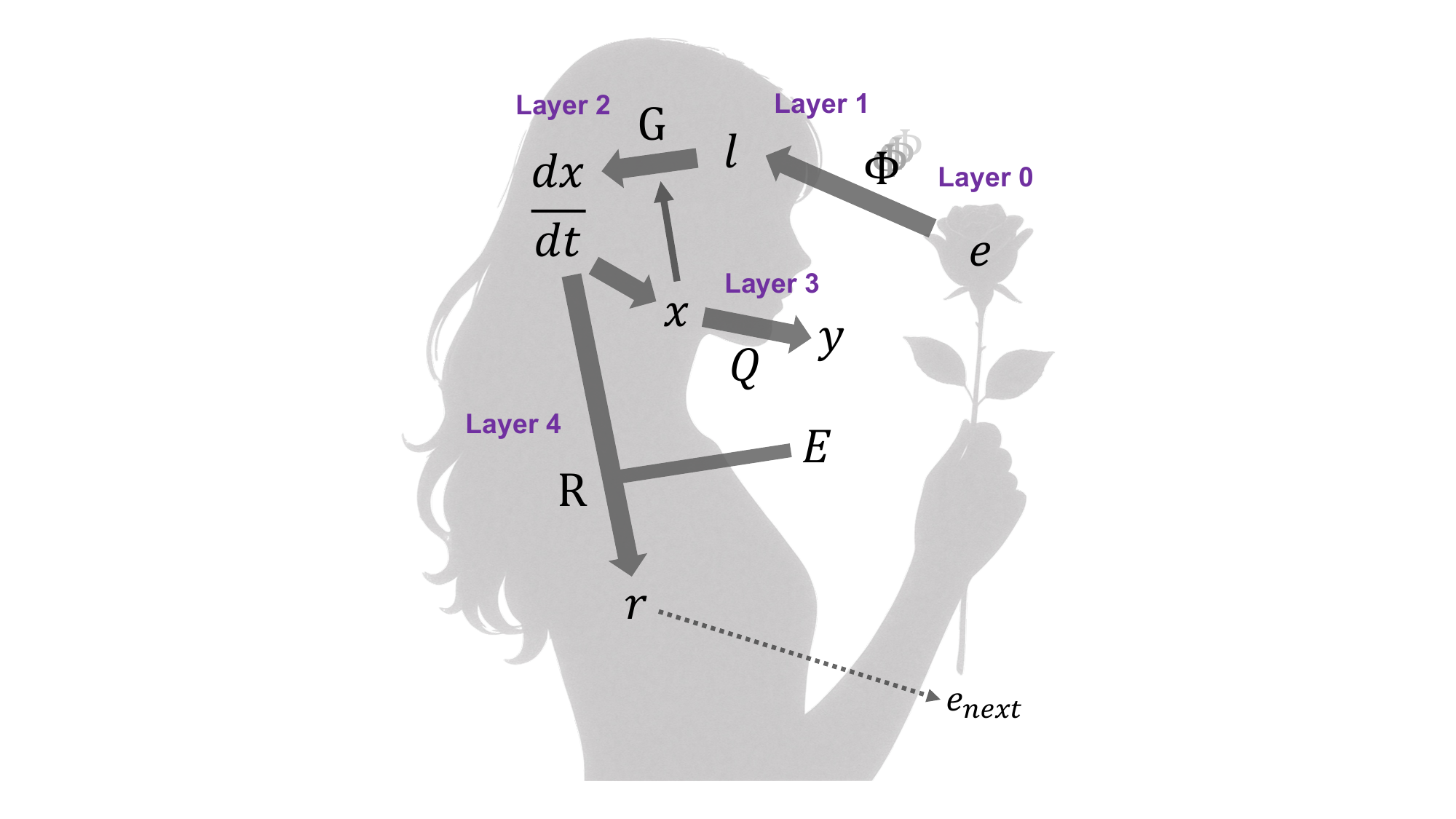}
  \caption{Five-layer descriptive framework for learning dynamics, illustrating the separation of descriptive responsibilities across external input, load generation, internal state transformation, observation, and subjective evaluation.}
  \label{fig:framework}
\end{figure}

The structural necessity of this design choice is examined in Appendix~\ref{app:structural-stability-eval-repr},
where it is shown that alternative couplings---most notably strictly state-dependent evaluation---generically induce
collapse of exploration and loss of sustained learning through the geometry of the evaluation--exploration loop itself.
The present formulation $r = R\!\left(\frac{dx}{dt}, E\right)$ thus arises not from normative assumptions about learning
objectives, but from structural constraints on the stability of adaptive dynamics.

%% file: sections/s04_layer_responsibilities.tex
\section{Summary of Layer Responsibilities}

\begin{table}[ht]
\centering
\begin{tabular}{ll}
\hline
\textbf{Layer} & \textbf{Descriptive Responsibility} \\
\hline
Layer~0 & External input \\
Layer~1 & Geometric decomposition \\
Layer~2 & Dynamic transformation \\
Layer~3 & Externalization (observation) \\
Layer~4 & Subjective evaluation (loop modulation) \\
\hline
\end{tabular}
\caption{Summary of descriptive responsibilities assigned to each layer of the framework.}
\label{tab:layer-responsibilities}
\end{table}

Each layer is defined by a single descriptive responsibility. The table functions as a structural index rather than a process diagram, summarizing what each layer is permitted to describe, rather than how learning unfolds over time.

The framework is explicitly designed to prevent the conflation of distinct explanatory roles, such as task structure, internal state change, observation, and evaluation, by assigning each to a separate layer with non-overlapping scope.

All layers are formulated in purely geometric and dynamical terms. No explicit value judgments, optimality criteria, or normative assumptions about learning outcomes are introduced at the level of the framework itself.

Individual differences in learning behavior or cognitive style are not represented as explicit state variables. Instead, such differences are expressed implicitly through the geometry, sensitivity, and noise characteristics of the dynamics and interfaces, as well as through the conditions under which particular mappings become operative, rather than as explicit state variables. This anti-attributional design preserves descriptive generality while allowing learner-dependent variation to be expressed structurally rather than as fixed traits.

%% file: sections/s05_positioning.tex
\section{Positioning}

The proposed five-layer framework does not advance a specific learning theory, nor does it aim to adjudicate among existing theoretical accounts. Instead, it is intended as a higher-level descriptive language in which established theories---such as Cognitive Load Theory---can be expressed as constrained or partial instances.

In this sense, the framework functions not as a standalone explanatory model, but as a formal descriptive substrate for comparing, relating, and extending different theories and observations at a shared structural level. Its role is not to determine which theory is correct, but to clarify what kinds of descriptions of learning processes are structurally admissible and how they may be coherently related.

%% file: sections/s06_conclusions.tex
\section{Conclusions}

This paper presented a five-layer formal descriptive framework for learning dynamics, designed as a symbolic language rather than a predictive or prescriptive model. By introducing explicit state variables, mappings, and layer-specific descriptive responsibilities, the framework provides a structurally transparent way to describe learning processes without reducing them to task difficulty, learner traits, or optimization objectives.

A central contribution of the framework is the explicit separation of descriptive responsibilities across external input, load generation, internal understanding transformation, observation, and subjective evaluation. This separation clarifies where different learning theories intervene within the learning process and avoids common forms of conceptual conflation, such as treating cognitive load as an intrinsic property of instructional materials or explaining persistence and withdrawal solely in terms of fixed motivational traits.

In particular, the introduction of a subjective evaluation interface allows learning persistence, stagnation, and avoidance to be described as adaptive control responses emerging from interactions between internal dynamics and environmental conditions. Within this framework, neither sustained learning nor avoidance is treated as inherently correct; instead, both are understood as context-dependent outcomes that may reflect biologically and structurally coherent regulation.

The proposed framework does not aim to replace existing learning theories. Rather, it provides a common descriptive language within which diverse theoretical perspectives can be expressed, compared, and extended without requiring premature formalization or commitment to a single explanatory principle. By prioritizing descriptive clarity and extensibility, the framework seeks to support cumulative understanding across disciplines and to serve as a foundation for future empirical and formal developments.

%% file: sections/s07_future_work.tex
\section{Future Work}

The present framework is intentionally formulated as a minimal descriptive language for learning dynamics, prioritizing structural clarity and extensibility over completeness. Several important extensions are therefore left for future work.

Most notably, the internal structure of the subjective evaluation interface $R$ is treated here as an abstract regulatory interface responding to learning dynamics, without commitment to specific reward semantics or motivational constructs. The analysis in Appendix~\ref{app:structural-stability-eval-repr} establishes wiring-level constraints on evaluation; a distinct open question is how the internal structure of $R$ realizes multiple modes and timescales while respecting those constraints. Empirical and theoretical considerations suggest that such evaluation may admit multiple control modes with distinct temporal and functional characteristics—such as sensitivity to change, sensitivity to accumulated change, and attenuation under sustained uncertainty. Articulating these modes remains a task for future work, to be pursued in a manner that preserves the descriptive and non-normative character of the framework.

A second direction concerns the treatment of external input $e$. In the current formulation, $e$ enters the understanding dynamics through load generation and state-dependent transformation, while its internal semantic or relational structure remains deliberately unspecified. Future extensions may introduce additional descriptive layers or mappings to characterize structural organization, temporal coherence, or relational dependencies within learning materials, as well as interactions among different load components generated by the same input, without collapsing these distinctions into a single representational level.

Finally, while the framework is presented as a descriptive language rather than a predictive model, its components are in principle amenable to empirical grounding. Future work will explore how variables and mappings in the framework can be operationalized through behavioral data, learning analytics, or experimental paradigms. Such efforts are not intended to replace the descriptive role of the framework, but to refine its expressive power as a unifying language for comparing learning environments, instructional designs, and adaptive systems.

More broadly, the framework opens the possibility of extending the same descriptive vocabulary across scales, from individual-level learning dynamics to population-level considerations. Variability in evaluation modes and control regimes may then be described as contributing to robustness and adaptability under changing environments, suggesting connections to biological perspectives on diversity, exploration--exploitation trade-offs, and long-term system resilience. These connections remain speculative at present but indicate promising directions for future descriptive and theoretical development.

%% file: appendices/appA_structural_stability.tex

\appendix
\section{Structural Stability Constraints on Evaluation and Representation Selection}
\label{app:structural-stability-eval-repr}

In this appendix, we analyze how differences in the coupling among evaluation, exploration, and internal dynamics affect the stability of learning, thereby clarifying the structural constraints underlying the evaluation interface introduced in Section~3.5.

Here, we consider an adaptive learning system in which the continuous change of the internal state $x$, $\frac{dx}{dt}$, and the interaction with the environment $E$ are controlled by an input acquisition policy. The evaluation $R$ is, in principle, a two-argument evaluation interface that depends on change in the internal state and feedback from the environment, thereby determining the evaluation signal $r$:
\[
r = R\!\left(\frac{dx}{dt}, E\right).
\]

In this setting, policy update and input acquisition may occur as discrete acquisition events indexed by $k \rightarrow k+1$:
\[
r_k
\xrightarrow{U} \pi_{k+1}
\xrightarrow{S} e_{k+1}.
\]

Here, $\pi$ is the input acquisition policy. $U$ denotes the policy update ($r_k \mapsto \pi_{k+1}$). $S$ denotes the input acquisition (sampling) operator ($\pi_{k+1} \mapsto e_{k+1}$).

The internal dynamics $\frac{dx}{dt}$ evolve continuously, and discrete acquisition events function as triggers that shape the subsequent continuous change. However, in order to visualize the events occurring in each loop succinctly, in this appendix we schematically approximate changes in the internal dynamics as discrete. Accordingly, we describe the change in $\frac{dx}{dt}$ mediated across steps by the internally driven loop as $\dot{x}_k \rightarrow \dot{x}_{k+1}$. Here, $\dot{x}_k$ is used as a representative symbol for $\frac{dx}{dt}$ within each step.

In what follows, we describe the same event by decomposing it into two schematics: a projection onto an internally driven loop and a projection onto a loop that includes interaction with the environment. The internally driven loop proceeds as follows:
\[
\dot{x}_k
\xrightarrow{R} r_k
\xrightarrow{U} \pi_{k+1}
\xrightarrow{S} e_{k+1}
\xrightarrow{\Phi} (\ell_{x,k+1}, \ell_{n,k+1})
\xrightarrow{G} \dot{x}_{k+1}.
\]

In addition, environmental feedback is generated through externalization and enters evaluation:
\[
x_k
\xrightarrow{Q} y_k
\xrightarrow{H} E_k
\xrightarrow{R} r_k
\xrightarrow{U} \pi_{k+1}
\xrightarrow{S} e_{k+1}
\xrightarrow{\Phi} (\ell_{x,k+1}, \ell_{n,k+1})
\xrightarrow{G} \dot{x}_{k+1}
\xrightarrow{\mathcal{I}} x_{k+1}.
\]

$\Phi$ is a fixed family of representation decomposition mappings. $Q$ denotes externalization ($x_k \mapsto y_k$). $H$ denotes the environmental response mapping ($y_k \mapsto E_k$). $G$ is a mapping that determines the internal dynamics (change in the internal state). $\mathcal{I}$ denotes the step-wise state update induced by the continuous-time flow (integration over the step).

In this sense, a single event may admit simultaneous projections of its components as $E$ and $e$ onto multiple functional bases (e.g., evaluation $R$ and acquisition $S$), without implying any serial transformation between them.

\subsection{Description as Probabilistic Sampling of $\Phi$-Selection}
\label{app:probabilistic-phi-selection}

In this appendix, we describe the selection of $\Phi$ as a process in which the policy $\pi$ acquires an input $e$, a bias over candidate $\Phi$ arises through the response to that $e$, and which $\Phi$ is chosen is then determined probabilistically according to the resulting weights. The probabilistic character here does not assume that an explicit probability distribution is represented inside the learner; rather, it is a descriptive summary of tendencies observed through repeated interaction. Furthermore, as long as the same $e$ is maintained for some period of time, $\Phi$-selection repeatedly occurs for that same $e$: in some cases the same $\Phi$ may continue to be selected, while in other cases different $\Phi$ may be selected. Accordingly, the gentler (less sharply peaked) the bias is, the more diverse $\Phi$ can be selected, making it possible to interpret a single $e$ from multiple perspectives; conversely, the sharper (more peaked) the bias is, the more likely the same $\Phi$ will continue to be selected, and deepening of understanding within the same context will occur. Moreover, this bias is not fixed: it may change over the long term through the accumulation of interactions with $e$ (in particular, through repeated contact with $e$ derived from others' $\Phi$). However, in this paper we treat such long-term changes in bias not as ``learning'' but as plastic changes in $\Phi$-selection (``noticing''), and we distinguish them from deepening of the internal state $x$ (``understanding'').

\subsection{Structural Degeneracy of State-Dependent Evaluation}
\label{app:degeneracy-state-dependent}

In this framework, we do not assume that evaluation takes the internal state $x$ as a direct input. This is because the internal state $x$ has no intrinsic reference scale by which it can be evaluated, even from the learner's own perspective. Without a standard of comparison, the absolute magnitude of $x$ cannot be internally monitored, whereas the internal change $\dot{x}$ may be experienced as variation over time. Moreover, coupling evaluation directly to $x$ risks conflating state description with control objectives. To make this structural consequence explicit, we temporarily consider the counterfactual wiring $r = R(x)$ and examine the resulting loop geometry. In discrete acquisition events, this is instantiated as $r_k = R(x_k)$.

\subsubsection{State-Dependent Evaluation (Strictly Internal Form)}
\label{app:state-dependent-internal-form}

We consider a strictly internal configuration in which evaluation is defined solely as a function of the internal state:
\[
r = R(x).
\]

Since the internal state $x$ is defined only relative to a chosen representational decomposition $\Phi$ and has no privileged intrinsic reference scale, state-based evaluation is not invariant under representational change. Consequently, the control objective induced by $R(x)$ is structurally tied not to any external criterion of success but to the maintenance of internal coherence under the currently active representation.

Under this wiring, the learning loop closes through evaluation, policy update, and acquisition, feeding back to internal variation.

\subsubsection{Structural Assumptions Underlying Degeneracy of State-Dependent Evaluation}
\label{app:assumptions-degeneracy}

The degeneracy of learning under strictly state-dependent evaluation arises not from any particular optimization objective but from the geometry of coupling among evaluation, policy update, and internal dynamics. The following mild structural assumptions suffice.

\paragraph{Assumption 1 (Evaluation-Preserving Policy Update).}
The policy update operator $U$ biases the input acquisition policy $\pi$ in the direction that makes it more likely for interaction patterns under which the evaluation signal $r$ remains high to occur again.

\paragraph{Assumption 2 (Structural Bias of State-Based Evaluation).}
When evaluation is defined as $r = R(x)$, the absence of an external reference structurally favors high values of $r$ in regions where internal representations are maintained as coherent and predictive under the currently active decomposition $\Phi$. Unless additional structure is artificially imposed on $R$, state-based evaluation therefore tends to assign persistently high $r$ in the direction that keeps change in the internal state small. This degeneracy can be avoided only by imposing additional structure that defines an evaluation scale invariant under representational changes (which is precisely what the present framework avoids assuming). Accordingly, the collapse described here is not tied to a particular objective, but follows generically from the coupling geometry when evaluation depends strictly on $x$.

\paragraph{Assumption 3 (Controllability of Internal Variation).}
Through the composition of the acquisition operator $S$, the representation family $\Phi$, and the internal dynamics $G$, the policy $\pi$ can adjust how the internal state changes by appropriately selecting the input $e$.

Under these conditions, the closed loop
\[
x_k \xrightarrow{R} r_k \xrightarrow{U} \pi_{k+1} \xrightarrow{S} e_{k+1} \xrightarrow{\Phi} (\ell_{x,k+1}, \ell_{n,k+1}) \xrightarrow{G} \dot{x}_{k+1}
\]
has a structurally stable attractor in the direction that keeps change in the internal state small.

The input acquisition policy $\pi$ converges toward input patterns that maintain persistently high values of the evaluation signal $r$ by suppressing change in the internal state. As a result, a stable behavior emerges in which change in the internal state is suppressed, but learning is degenerate: internal representations remain coherent under the current decomposition $\Phi$, while the system loses the capacity for sustained adaptation.

\subsection{Transition-Dependent Evaluation}
\label{app:transition-dependent-evaluation}

To avoid the structural collapse induced by state-dependent evaluation, the evaluation interface must be coupled to quantities that are robust to representational change and that reflect actual interaction with the environment. This requirement is satisfied when evaluation is defined as:
\[
r = R\!\left(\frac{dx}{dt}, E\right).
\]
In discrete acquisition events, we write this schematically as $r_k = R(\dot{x}_k, E_k)$.

Under this coupling, high values of $r$ cannot be maintained merely by suppressing internal change. Internal state $x$ becomes evaluable only insofar as it is consistent with external consequences $E$, and conversely, environmental feedback becomes meaningful only relative to the internal transformations that produced it.

As a result, the learning loop acquires a fundamentally different geometry:
\[
x_k \xrightarrow{Q} y_k \xrightarrow{H} E_k \xrightarrow{R} r_k \xrightarrow{U} \pi_{k+1} \xrightarrow{S} e_{k+1} \xrightarrow{\Phi} (\ell_{x,k+1}, \ell_{n,k+1}) \xrightarrow{G} \dot{x}_{k+1} \xrightarrow{\mathcal{I}} x_{k+1}.
\]

In this configuration, because evaluation constrains internal dynamics through externally grounded consequences, the policy $\pi$ cannot preserve high values of $r$ by suppressing change in the internal state. Therefore, exploration through $S$ persists, the effective distribution over representation decompositions $\Phi$ maintains diversity, and long-term adaptation becomes structurally stable.

When environmental feedback $E$ becomes persistently adverse, the update $U$ may favor acquisition policies that prioritize immediate reduction of adverse consequences, yielding a stabilization regime (a short-term protective mode) in which inputs are sampled to stabilize the $E$-channel rather than to diversify $\Phi$-selection. In such a stabilization regime, sampling tends to concentrate on a narrower subset of interaction patterns (even when the nominal input class $e$ is unchanged), which can sharpen the effective bias over $\Phi$ and make alternative decompositions less likely. Note that long-term sharpening of the effective bias in $\Phi$-selection may arise either from plastic changes in the evaluation interface $R$ (e.g., shifts in sensitivity or gain with respect to $(\dot{x}, E)$) or from changes in the effective environmental feedback channel (e.g., via $x \xrightarrow{Q} y \xrightarrow{H} E$); the present framework is compatible with both interpretations.

This structural distinction is independent of any particular optimization objective or learning algorithm. It is a consequence of the coupling geometry among evaluation, exploration, and internal dynamics.

\subsection{Conclusion}
\label{app:conclusion}

The stability of learning in adaptive systems is determined not by the form of optimization objectives or the details of learning algorithms but by the coupling geometry among evaluation, exploration, and internal dynamics.

When evaluation is coupled directly to the internal state, $r = R(x)$, the learning loop generically acquires a structurally stable attractor in which internal variation is suppressed, exploration collapses, and sustained learning attenuates. This degeneration arises not from a particular design choice but from the representational dependence of state itself. 

By contrast, when evaluation is defined over internal transition and environmental feedback, $r = R\!\left(\frac{dx}{dt}, E\right)$, the learning loop is structurally constrained so as to preserve exploration, maintain representational diversity, and support long-term adaptation. In this configuration, learning stability emerges as a geometric property of the evaluation--exploration coupling. 

Thus, the persistence of learning is governed not by what is optimized but by how the system is wired. In this sense, this framework characterizes learning not as objective maximization but as a problem of structural coupling. The degeneracy result holds for any learning architecture satisfying Assumptions~1--3, independent of parameterization or learning rule.

%% file: sections/s08_acknowledgements.tex
This work was developed through iterative reasoning with large language models, including ChatGPT (OpenAI) and Claude (Anthropic). These AI systems were used as intellectual tools to support the research process, primarily by assisting with articulation, clarification, and structured exploration of ideas, including the refinement of formal definitions, the identification of logical inconsistencies, and the examination of possible implications of the framework.

All core theoretical structures, design decisions, and conceptual contributions were conceived, evaluated, and finalized by the human author. The AI systems did not act as independent agents or co-authors, but functioned as instruments for externalizing, reflecting on, and stress-testing the author’s reasoning during the development of the work.

The author also wishes to express gratitude to the many individuals---including family members, colleagues, students, and collaborators---whose interactions, discussions, and shared experiences over the years provided essential experiential input beyond the scope of formal modeling, and formed part of the broader context from which this framework emerged.